\documentclass[12pt, a4paper,superscriptaddress,aps,nofootinbib, showkeys, showpacs, ]{revtex4}
\usepackage{amsmath,amssymb,graphicx}
\makeatletter
\usepackage[active]{srcltx}
\usepackage{graphicx,color}
\usepackage{changebar}
\usepackage{hyperref}
\usepackage[T1]{fontenc}
\usepackage{ae}
\usepackage[ansinew]{inputenc}
\usepackage{amsmath}
\usepackage{braket}
\usepackage{mathtools}
\usepackage{slashed}
\usepackage{empheq}
\usepackage{tikz}
\usepackage{multirow}
\usetikzlibrary{decorations.pathmorphing}
\usepackage{amsmath,amssymb,graphics,epsfig,subfigure}
\usetikzlibrary{quotes,angles}
\usepackage{etoolbox}

\usepackage{graphicx}
\usepackage{amsfonts}
\usepackage{amsmath}
\usetikzlibrary{arrows}
\usetikzlibrary{decorations.markings}

\begin{document}

\title{Rotation effect for chiral magnetic cosmic string in the Kaluza-Klein theory}

\author{R. L. L. Vit\'oria} 
\email{ricardovitoria@professor.uema.br/ricardo-luis91@hotmail.com}
\affiliation{Faculdade de F\'isica, Universidade Federal do Par\'a, Av. Augusto Corr\^ea, Guam\'a, 66075-110, Bel\'em, PA, Brazil.}
\author{C. F. S. Pereira} 
\email{carlos.f.pereira@edu.ufes.br}
\affiliation{Departamento de F\'isica e Qu\'imica, Universidade Federal do Esp\'irito Santo, Av.Fernando Ferrari, 514, Goiabeiras, Vit\'oria, ES 29060-900, Brazil.}
\author{E. V. B. Leite} 
\email{ericovbleite@gmail.com}
\affiliation{Departamento de F\'isica, Universidade Federal da Para\'iba, Caixa Postal 5008, 58051-900 Jo\~ao Pessoa, PB, Brazil}
\author{A. R. Soares }
\email{adriano.soares@ifma.edu.br/adriano2da@gmail.com}
\affiliation{Instituto Federal de Educa\c{c}\~ao Ci\^encia e Tecnologia do Maranh\~ao, Campus Buriticupu, CEP 65393-000, Buriticupu, Maranh\~ao, Brazil.}
\author{H. Belich} 
\email{humberto.belich@ufes.br}
\affiliation{Departamento de F\'isica e Qu\'imica, Universidade Federal do Esp\'irito Santo, Av.Fernando Ferrari, 514, Goiabeiras, Vit\'oria, ES 29060-900, Brazil.}

\begin{abstract}
We have investigated bound state solutions for a scalar particle described in the spacetime of a chiral cosmic string subjected to rotational effects in a Kaluza-Klein theory. We saw that combinations between the parameters that describe the generalized spacetime impose upper and lower limits for the radial coordinate leading to well-established boundary conditions.

\end{abstract}


	
\maketitle	

\newpage

\section{Introduction}\label{sec1}

In the development of grand unification theories for fundamental interactions, the emergence of topological defects appears naturally in phase transition processes in a young universe \cite{INTRO1, INTRO2, INTRO3, INTRO4}. Among a variety of defect type models, we have the global monopole \cite{INTRO5,INTRO6,INTRO7}, domain wall \cite{INTRO3,INTRO8}, cosmic strings \cite{INTRO8,INTRO9,INTRO10,INTRO11,INTRO12}. Although we have no observational evidence for their existence, cosmic strings are candidates for possible observation \cite{INTRO3}. 

In particular, cosmic string solutions that are written in the context of general relativity are bridged with vortex solutions, which are displayed in flat spacetime. As vortex solutions are linked to the characteristics of superconductivity, it is possible to associate macroscopic with microscopic details \cite{INTRO12,2}. Still in the cosmic string scenario, this solution was investigated in the aspect of non-conservative theory \cite{INTRO15,INTRO16} and Born-Infeld theory \cite{INTRO17,INTRO18}. Similarly, the global monopole was investigated in the Born-Infeld theory and later some other effects were obtained \cite{INTRO19,INTRO20,INTRO21}.

Such ideas, involving topological defects, can be studied in an extra-dimensional setting through Kaluza-Klein theory (KKT). This theory stems from the ideas of Theodor Kaluza , in his paper released in 1921 \cite{KALUZA}, and Oskar Klein in 1926 \cite{KLEIN}, where the authors propose unifying the gravitation and electromagnetism, as curvature  effects of a pseudo-Riemannian manifold in five-dimensional spacetime. KKT was applied in different scenarios such as: in studies on graphene \cite{KK1}, in magnetic cosmic string \cite{KK2}, on a Dirac field \cite{KK3}, on a Klein-Gordon particle with the position-dependent-mass \cite{KK4} and on the Landau Levels \cite{KK5}.

The main motivation in this work is to investigate a chiral magnetic cosmic string in KKT, but under the rotation effect. Proposed by Landau and Lifshitz \cite{landau}, rotation effects in Minkowski spacetime with cylindrical symmetry serve to investigate two phenomena, one of which is our focus here: the singular behaviour at large distances for a system in a uniformly rotating frame of which imposed a restriction on the spatial coordinates due to the effects of rotation. Noninertial effects due to the rotation have been investigated in several quantum systems scenarios, for example, on scalar bosons \cite{R2}, on a scalar particle in cosmic string spacetime \cite{R3}, on a scalar field in the spacetime with space-like dislocation and in the spacetime with a spiral dislocation \cite{me}, on a scalar field in a spacetime with a magnetic
screw dislocation \cite{me1}, on a scalar field in a time-dislocation space-time \cite{KVR} and on a scalar field in a KKT \cite{R1}.

In particular, in Ref. \cite{KVR} we can see that the restriction on the spatial coordinates depends on time-like torsion, while in Ref. \cite{R1} the restriction on the spatial coordinates depends on the quantum flux coming from the extra dimension of KKT. In both cases, it is possible to observe the quantum effects on the bound state solutions of the proposed systems due to these restrictions modified by the background. However, there is still no study in the literature that involves an interface between extra dimension and time-type displacement. From a theoretical point of view, it is a possible scenario and, in addition, it is a generation of cases investigated separately.

The structure of this paper is as follows: in Sec. (\ref{sec2}), we carry out the necessary weightings and build a generalized space-time; in Sec. (\ref{sec3}), we analyze the general solution and construct the bound states for asymptotic configurations; finally, in Sec. (\ref{sec4}), we present our conclusions.

\section{The background}\label{sec2}

Here we will describe the spacetime we are interested in working with. Taking as a starting point the line element that describes the space-times of a magnetic cosmic string in KKT. The constant magnetic flux term is added to the extra dimension by representing an effect analogous to the Aharonov-Bohm effect (ABE) \cite{ab,1} whose quantum dynamics for a scalar particle was investigated in Ref. \cite{2}. Thus, the line element describing the space-time of KKT under the effect of an external field is defined by
\begin{eqnarray}\label{1}
ds^2= -(dt+Jd \varphi)^2 + d\rho^2 + \alpha^2\rho^2{d\varphi^2} + (dz+\chi d\varphi)^2+ \left(d\omega+ A{d\varphi}\right)^2,
\end{eqnarray}
where $\omega$ represents the coordinate of the extra dimension, $\alpha$ the angular defect of the cosmic string, $\chi$ and $J$ are the parameters associated to the like-space and like-time dislocations (torsion), respectively \cite{put} and $A$ is associated with the constant magnetic flux term. The angular defect $\alpha=1-4\mu$ is defined in terms of the string linear mass density. We are defining the potential vector $A_\nu=\left(0,0,A_{\varphi},0\right)$ whose non-zero component takes the form $A={\kappa}A_{\varphi}$  being $A_{\varphi}=\frac{\Phi}{2\pi\kappa}$, with $\Phi$ being the constant quantum flux.

As our interest is to investigate a generalized scenario of those studied in the references \cite{2,3,KVR}. Then, we will insert in the metric Eq. (\ref{1}) the characteristics of a dislocated magnetic cosmic string that is subjected to rotation effects. In this way, we will make the following transformation \cite{landau} in the line element Eq. (\ref{1}):
\begin{eqnarray}\label{2}
\varphi \longrightarrow \varphi + \Omega{t},
\end{eqnarray}
by resulting in the following line element
\begin{eqnarray}\label{3}
ds^2&=&- \left[1-\Omega^2\left(\alpha^2\rho^2-\zeta\right)\right]dt^2 + d\rho^2+d\omega^2+dz^2 + \left[\alpha^2\rho^2+\frac{2J}{\Omega}-\zeta\right]d\varphi^2 \\\nonumber
&+& 2A{d\omega}{d\varphi} + 2A\Omega{d\omega}{dt} + 2\chi{dz}{d\varphi}+2\chi\Omega{dz}{dt} + 2\Omega\left[\alpha^2\rho^2+\frac{J}{\Omega}-\zeta\right]d\varphi{dt},
\end{eqnarray}
with $\zeta=\frac{2J}{\Omega}+J^2-\left(A^2+\chi^2\right)$. In the limit $\Omega\longrightarrow{0}$ this line element describes the space-time of a dislocated chiral cosmic string, according to investigation in Ref. \cite{2}.

As we want to maintain the causal connection for objects described in this space-time, we need to analyze and ensure that the components $g_{tt}>0$ and $g_{\varphi\varphi}>0$ are positive. Such a requirement leads us to the following constraint for radial coordinates
\begin{eqnarray}\label{4}
\frac{\sqrt{J^2-\left(A^2+\chi^2\right)}}{\alpha}< \rho < \frac{\sqrt{\left(1+\Omega{J}\right)^2-\Omega^2\left(A^2+\chi^2\right)}}{\alpha\Omega}.
\end{eqnarray}
that is, unlike Minkowski spacetime, the axial coordinate now has a validity interval. This effect is due to non-inertial effects and time-like dislocation in spacetime. Furthermore, we can note that this range of validity depends on the parameters that characterize the non-triviality of the background, that is, the minimum and maximum limits are influenced by the cosmic string, the space-like and time-like displacements and the extra dimension.

We can note that, by taking $J=0$, $A=0$, $\chi=0$ and $\alpha=1$ into Eq. \ref{4} we recover the interval defined in Ref. \cite{landau}; by taking $J=0$, $A=0$, $\chi=0$ and $0<\alpha<1$ into Eq. (\ref{4}) we recover the interval defined in Refs. \cite{R2,R3,me}; by making $J$ non null, $A=0$, $\chi=0$ and $\alpha=1$ into Eq. (\ref{4}) we recover the result obtained in Ref. \cite{KVR}.

\section{Quantum Dynamics}\label{sec3}

This section will be reserved for investigating the quantum dynamics of a scalar particle in this generalized space-time defined in Eq.(\ref{3}). We begin then by writing the covariant form of the Klein-Gordon equation for a scalar field 
\begin{eqnarray}\label{5}
\frac{1}{\sqrt{-g}} \partial_{\mu}\left[\sqrt{-g}g^{\mu\nu}\partial_\nu\right]\phi -m^2\phi=0,
\end{eqnarray} 
where $g=det\left(g_{\mu\nu}\right)=-\alpha^2\rho^2$ and $g^{\mu\nu}$ is the inverse of the metric tensor which is defined by
\[
  g^{\mu\nu} =
  \left[ {\begin{array}{ccccc}
    -\left(1-\frac{J^2}{\alpha^2\rho^2}\right) & 0 & \Omega-\frac{J\left(1+\Omega{J}\right)}{\alpha^2\rho^2} & \frac{\chi{J}}{\alpha^2\rho^2} & \frac{AJ}{\alpha^2\rho^2} \\
   0 & 1 & 0 & 0 & 0\\
   \Omega-\frac{J\left(1+\Omega{J}\right)}{\alpha^2\rho^2}  & 0 & \frac{1}{\alpha^2\rho^2}-\Omega^2+ \frac{J\Omega\left(2+\Omega{J}\right)}{\alpha^2\rho^2} & -\frac{\chi\left(1+\Omega{J}\right)}{\alpha^2\rho^2} & -\frac{A\left(1+\Omega{J}\right)}{\alpha^2\rho^2}\\
   \frac{\chi{J}}{\alpha^2\rho^2} & 0 &  -\frac{\chi\left(1+\Omega{J}\right)}{\alpha^2\rho^2} & 1+\frac{\chi^2}{\alpha^2\rho^2} & \frac{A\chi}{\alpha^2\rho^2} \\
   \frac{A{J}}{\alpha^2\rho^2} & 0 &  -\frac{A\left(1+\Omega{J}\right)}{\alpha^2\rho^2} & \frac{A\chi}{\alpha^2\rho^2} & 1+ \frac{A^2}{\alpha^2\rho^2}
  \end{array} } \right].
\]

Therefore, by substituting Eq.(\ref{3}) into Eq.(\ref{5}) we obtain
\begin{eqnarray}\label{6}
&-&\left(1-\frac{J^2}{\alpha^2\rho^2}\right)\frac{\partial^2\phi}{\partial{t}^2} + \frac{\partial^2\phi}{\partial{\rho}^2} + \frac{1}{\rho}\frac{\partial\phi}{\partial{\rho}} + \left[\frac{1}{\alpha^2\rho^2}-\Omega^2+ \frac{J\Omega\left(2+\Omega{J}\right)}{\alpha^2\rho^2}\right]\frac{\partial^2\phi}{\partial{\varphi}^2}\\\nonumber
&+&\left(1+\frac{\chi^2}{\alpha^2\rho^2}\right)\frac{\partial^2\phi}{\partial{z}^2}+\left(1+\frac{A^2}{\alpha^2\rho^2}\right)\frac{\partial^2\phi}{\partial{\omega}^2}+ 2\left[\Omega-\frac{J\left(1+\Omega{J}\right)}{\alpha^2\rho^2}\right]\frac{\partial^2\phi}{\partial{t}\partial\varphi} + \left(\frac{2J\chi}{\alpha^2\rho^2}\right)\frac{\partial^2\phi}{\partial{t}\partial{z}} \\\nonumber
&+& \left(\frac{2JA}{\alpha^2\rho^2}\right)\frac{\partial^2\phi}{\partial{t}\partial{\omega}}- \left[\frac{2\chi\left(1+\Omega{J}\right)}{\alpha^2\rho^2}\right]\frac{\partial^2\phi}{\partial{\varphi}\partial{z}} -\left[\frac{2A\left(1+\Omega{J}\right)}{\alpha^2\rho^2}\right]\frac{\partial^2\phi}{\partial{\varphi}\partial{\omega}}- \left(\frac{2A\chi}{\alpha^2\rho^2}\right)\frac{\partial^2\phi}{\partial{z}\partial{\omega}} - m^2\phi=0
\end{eqnarray} 

We can observe that in limits of $\chi\to{0}$ and $J\to{0}$ we recover the same differential equation obtained in Ref. \cite{3}.

The line element that describes the spacetime given in Eq. (\ref{1}) has axial symmetry. Thus, with the objective of separating variables, we propose the following ansatz as a solution to the general differential equation (\ref{6})
\begin{eqnarray}\label{7}
\phi\left(t,\rho,\varphi,z,\omega\right)= f\left(\rho\right)e^{-i\left(Et-\varphi{l}-kz-q\omega\right)},
\end{eqnarray}	where $l= 0, \pm{1}, \pm{2}$, . . . is eigenvalue of angular momentum operator $\hat{L_z}=-i\partial_\varphi$, $-\infty< k <\infty$ is eigenvalue linear moment operator $\hat{p_z}=-i\partial_{z}$ and $q=const$. is the eigenvalue of the linear momentum operator that is associated with the extra dimension $\hat{p_\omega}=-i\partial_{\omega}$. Finally, $E$ is a constant that is linked to the energy eigenvalue and $f\left(\rho\right)$ is a function associated with the axial coordinate. Then, by substituting Eq. (\ref{7}) into Eq. (\ref{6}), we find a Bessel equation \cite{livro1,livro2}
\begin{eqnarray}\label{8}
{f}''+\frac{{f}'}{\rho} -\frac{\gamma^2}{\rho^2}f+ \beta^2f =0,
\end{eqnarray} 
where the parameters $\gamma^2$ and $\beta^2$ are defined as
\begin{eqnarray}\label{9}
\beta^2&=& \left(E+l\Omega\right)^2-k^2-q^2-m^2, \\\nonumber
\gamma^2&=& \frac{1}{\alpha^2}\left[E^2J^2 +\chi^2k^2+ \left(l-qA\right)^2+ J\Omega{l^2}\left(2+\Omega{J}\right) + 2lJ\left(1+\Omega{J}\right)E\right] \\\nonumber
&-&\frac{1}{\alpha^2}\left[2k\chi{JE}+2qA{JE}+2lk\chi\left(1+\Omega{J}\right)+2lqA\Omega{J}+2Aqk\chi\right].
\end{eqnarray}

The solution of the above differential equation given in Eq. (\ref{8}) is given in terms of Bessel functions of the first $J_{|\gamma|}$ and second $N_{|\gamma|}$ kinds
\begin{eqnarray}\label{10}
f\left(\rho\right)= C J_{|\gamma|}\left(\beta\rho\right) + D N_{|\gamma|}\left(\beta\rho\right),
\end{eqnarray} 
for $C$ and $D$ being constants. 

Based on the radial constraint defined in Eq. (\ref{4}) due to the requirement to write objects that lie inside the light cone. We can adopt the minimum and maximum limits of the axial coordinate imposed by the time-like dislocation and rotation, respectively, as concentric cylindrical barriers, as adopted in Ref. \cite{KVR}. That is, will have to condition that our wave function is zero in these limits
\begin{eqnarray}\label{11}
f\left(r_1= \frac{\sqrt{J^2-\left(A^2+\chi^2\right)}}{\alpha}\right)=0, \qquad  f\left(r_2= \frac{\sqrt{\left(1+\Omega{J}\right)^2-\Omega^2\left(A^2+\chi^2\right)}}{\Omega\alpha}\right)=0.
\end{eqnarray}
Thus, applying these edge conditions Eq. (\ref{11}) in the general solution of Eq. (\ref{10}), we obtain the following equation that must be satisfied in these contours
\begin{eqnarray}\label{12}
N_{|\gamma|}\left(\beta{r_2}\right)J_{|\gamma|}\left(\beta{r_1}\right)-N_{|\gamma|}\left(\beta{r_1}\right)J_{|\gamma|}\left(\beta{r_2}\right)=0.
\end{eqnarray} 

Note that the above equation does not depend on the constants $C$ and $D$. Even though it is not possible to obtain bound states for this combination of Bessel functions of the first and second kinds Eq. (\ref{12}) due to origin regularity problems, we can work with its asymptotic version. For this, let's consider that $\beta{r_j}>>1$ and $|\gamma|$ is fixed. Then, we can perform the expansion of Bessel functions in terms of Hankel functions \cite{livro1,4,5}. So,
\begin{eqnarray}\label{13}
J_{|\gamma|}\left(\beta{r_j}\right)\approx \sqrt{\frac{2}{\pi{\beta{r_j}}}}\left[\cos\left(\beta{r_j}-\frac{\pi}{4}-\frac{\gamma\pi}{2}\right)-\left(\frac{4\gamma^2-1}{8\beta{r_j}}\right)\sin\left(\beta{r_j}-\frac{\pi}{4}-\frac{\gamma\pi}{2}\right)\right], \\\nonumber
N_{|\gamma|}\left(\beta{r_j}\right)\approx \sqrt{\frac{2}{\pi{\beta{r_j}}}}\left[\sin\left(\beta{r_j}-\frac{\pi}{4}-\frac{\gamma\pi}{2}\right)+\left(\frac{4\gamma^2-1}{8\beta{r_j}}\right)\cos\left(\beta{r_j}-\frac{\pi}{4}-\frac{\gamma\pi}{2}\right)\right],
\end{eqnarray}
where $j=1,2$. Therefore, replacing Eq. (\ref{13}) in Eq. (\ref{12}), approximating small angles and keeping only terms from up to $\mathcal{O}\left(\frac{1}{r^2}\right)$ we find the expression of the approximate energy spectrum

\begin{eqnarray}\label{14}
\beta^2 \approx \frac{\left(4\gamma^2-1\right)}{4r_1r_2} + \frac{n^2\pi^2}{\left(r_2-r_1\right)^2}
\end{eqnarray} with $n=0,1,2,3,4$, . . . the quantum number by representing the radial mode.

Before we set out to write the energy spectrum of in its explicit form, we need to direct our gaze to Eq. (\ref{9}) which has terms that depend on the constant $E$ and then perform some redefinitions. So, we have to $\gamma^2={\gamma^2_1}+ {\gamma^2_2}E+ \frac{E^2J^2}{\alpha^2}$ where
\begin{eqnarray}\label{15}
{\gamma^2_1}&=& \frac{1}{\alpha^2}\left[k^2\chi^2+Jl^2\Omega\left(2+\Omega{J}\right)-2lk\chi\left(1+\Omega{J}\right)-2lqA\Omega{J}-2Akq\chi+\left(l-qA\right)^2\right], \\\nonumber
{\gamma^2_2}&=& \frac{1}{\alpha^2}\left[2Jl\left(1+\Omega{J}\right)-2kJ\chi-2qAJ\right].
\end{eqnarray}

Therefore, replacing the definitions contained in Eq. (\ref{9}) in the approximate spectrum expression Eq. (\ref{14}) and using the redefinitions Eq. (\ref{15}). We were able to obtain a quadratic algebraic equation for the constant that is associated with the energy eigenvalue
\begin{eqnarray}\label{16}
{\gamma^2_3}E^2 + {\gamma^2_4}E+ P =0,
\end{eqnarray} 
where the above constants are defined as
\begin{eqnarray}\label{17}
{\gamma^2_3}&=& \left(1- \frac{J^2}{\alpha^2{r_1r_2}}\right), \qquad \qquad {\gamma^2_4}= \left(2l\Omega-\frac{{\gamma^2_2}}{r_1r_2}\right), \\\nonumber
P&=& \left(\frac{1}{4r_1r_2}\right)-\frac{n^2\pi^2}{\left(r_2-r_1\right)^2} + \Omega^2{l^2}- \frac{{\gamma^2_1}}{r_1r_2}-k^2-q^2-m^2.
\end{eqnarray}

Finally, we can write the approximate energy spectrum that contains all radial modes
\begin{eqnarray}\label{18}
E&\approx& \frac{\left[Jl\left(1+\Omega{J}\right)-kJ\chi-qAJ-\Omega{l\alpha^2}r_1r_2\right]}{\left(\alpha^2{r_1r_2}-J^2\right)} \\\nonumber
&\pm & \frac{\left[\left(2l\Omega-\frac{\gamma^2_2}{r_1r_2}\right)^2+4\left(1-\frac{J^2}{\alpha^2{r_1r_2}}\right)\left(k^2+q^2+m^2-l^2\Omega^2+\frac{\gamma^2_1}{r_1r_2}+\frac{n^2\pi^2}{\left(r_2-r_1\right)^2}-\frac{1}{4r_1r_2}\right)\right]^\frac{1}{2}}{2\left(1-\frac{J^2}{\alpha^2{r_1r_2}}\right)}.
\end{eqnarray}

In the energy spectrum expression above Eq. (\ref{18}) was dependent on $r_1,r_2, \gamma^2_1$ and $\gamma^2_2$. Where the constants $\gamma^2_1$ and $\gamma^2_2$ are defined in Eq. (\ref{15}) and $r_1,r_2$ defined in Eq. (\ref{11}). Thus, we have the energy spectrum that describes all allowed energy levels for the referred radial modes. Such configurations represent the formation of bound states for a massive scalar particle described in a generalized space-time. It is noteworthy that this discrete energy spectrum for the scalar field Eq. (\ref{18}) is only possible due to restrictions imposed on the internal $r_1$ and external $r_2$ radius of the cylindrical surface. Thus establishing the boundary conditions for the radial function.
 
In particular, we can consider the limits associated with the extra dimension $q$, the space-like dislocation $\chi$, cosmic string $\alpha$, the flux term $A$ all going to zero that we recover the same energy spectrum investigated in Ref. \cite{KVR}
\begin{eqnarray}\label{19}
E_{n,l,k}\approx \pm\sqrt{\left(k^2+m^2\right)\left(1+\Omega{J}\right)+ \Omega^2\pi^2\alpha^2{n^2}\left(1+\Omega{J}\right)+l^2\Omega^2+ \Omega^2\left(\frac{l^2}{J\Omega}-\frac{\alpha^2}{4J\Omega}\right)}.
\end{eqnarray}

In particular, we can consider the limits associated with the parameter that multiplies the extra dimension $q$, the space-like dislocation $\chi$, the cosmic string $\alpha$ and the quantum flux term $A$ all going to zero that we recover the same energy spectrum investigated in Ref. \cite{KVR} less than the coefficient that characterizes the defect of the cosmic string $\alpha$. On the other hand, if we consider that these radius $r_1$ and $r_2$ represent two concentric cylinders the restriction to the inner radius makes it physically not possible to try to return to consistent results in Ref. \cite{3}. This happens because we need to consider the parameters $J$ and $\chi$ to the zero. Thus, observe in the constraint of the inner radius $\sqrt{J^2-\chi^2-A^2}$ that in these limits the constant associated with the magnetic flux $A$ becomes imaginary. This leads us to interpret that when constructing a more general metric, the coefficients linked to the time-like dislocation must be greater than the sum of the magnetic flux parameters and the radial dislocation $J^2>A^2+\chi^2$.

\section{Conclusions}\label{sec4}

We have investigated the effects of rotation on a metric that falls into a non-trivial background, which is characterized by the presence of twist and curvature. The existing torsion is of two types: space-type dislocation and time-type dislocation. The curvature is characterized by the presence of the cosmic string. In addition to these ingredients, we insert an extra dimension into space-time through a KKT, through the presence of a constant quantum flow analogous to ABE.

Naturally, the more effects that are inserted at the bottom of the defect, the stronger are the restrictions that arise between the parameters that characterize each scenario. Such restrictions arise naturally when the requirement is made to maintain the causal structure of objects described in this generalized spacetime, making all events occur only within the light cone. Hence, the radial coordinate becomes bounded from above and below.

The general solution of the equation of motion is given by Bessel functions of the first and second kinds, which when imposed on boundary conditions, we are left with an expression that is defined by combinations of the Bessel functions on the boundaries.
The general solution of the equation of motion is given by Bessel functions of the first and second kinds, which when imposed on boundary conditions, we are left with an expression that is defined by combinations of the Bessel functions on the boundaries.

Finally, we analyzed the possibility of recovering the results obtained in other studies carried out in the literature. As well as the case studied in Ref. \cite{KVR} for a scalar particle in a spacetime with uniform rotation and a temporal dislocation. Within the due limits imposed on the general expression of the energy spectrum Eq. (\ref{18}), the result of \cite{KVR} less than the parameter of the topological defect $\alpha$ is recovered Eq. (\ref{19}).
	
\begin{acknowledgments}
The authors would like to thank CNPq and CAPES Brazilian Agencies, for financial support.
\end{acknowledgments}
	
\section*{Data Availability}
	
Data Availability Statement This manuscript has no associated data or the data will not be deposited. (Authors' comment: This is a theoretical study and no experimental data has been listed.)


\end{document}